# Characterization of high current pulsed arcs ranging from 100 kA to 250 kA peak

**R Sousa Martins[1], C Zaepffel[1], L Chemartin[1], Ph Lalande[1] and F Lago[2]**

[1]DPHY, ONERA, Université Paris Saclay, F-91123 Palaiseau – France
[2]DGA Techniques aéronautiques, 47 rue Saint-Jean, BP 53123, 31131 Balma Cedex – France

E-mail: rafael.sousa_martins@onera.fr

**Abstract**

In this paper, we present the study realized on three experimental setups that produce in laboratory a free arc channel subjected to the transient phase of a lightning current waveform. This work extends the high current pulsed arc characterization performed on previous studies for peak levels up to 100 kA. Eleven high current waveforms with peak value ranging from 100 kA to 250 kA with different growth rates and action integrals are studied, allowing the comparison of different test benches. These waveforms correspond to standard lightning ones used in aircraft certification processes. Hydrodynamic properties such as arc channel evolution and shock wave propagation are determined by high speed video imaging and Background-Oriented Schlieren method. The arc diameter reaches around 90 mm at 50 µs for a current of 250 kA peak. Space- and time-resolved measurements of temperature, electron density and pressure are assessed by optical emission spectroscopy associated with the radiative transfer equation. It is solved across the arc column and takes into account the assumption of non-optically thin plasma at local thermodynamic equilibrium. For a 250 kA waveform, temperatures up to 43000 K are found, with pressures in the order of 50 bar. The influence of current waveform parameters on the arc properties are analysed and discussed.

## 1. Introduction

Aeronautics is highly concerned about lightning strike, since an aircraft can expect, on average, one lightning strike every 1000 to 10000 flight hours [1, 2]. It is a possible safety hazard for aircraft, imposing thermo-mechanical constraints to the aircraft structure and also electromagnetic coupling with internal cables and components, due to the high current circulation. The concerns are even more significant for recent airline programs, which employ carbon fibre composite in around 50% of the aircraft weight. Carbon fibre composites are more vulnerable to lightning effects as a result of their lower thermal and electrical conductivities when compared to metallic structures. For that reason, important steps of lightning protection specifications and certification processes must be realized [3]. Those processes are accomplished mainly through lightning laboratory tests, which deal with high current waveforms up to 200 kA peak [4].

Aiming to optimize the aeronautical protections and to reduce the need to perform numerous lightning tests for certification, reliable computational tools able to better predict the behaviour of lightning arcs are more and more required. These simulation codes request inputs and validations taken from experimental databases. However, sufficient and accurate data is yet missing for high current arcs. The lack of understanding is principally related to the transient phase of the lightning arc, which can reach thousands of amperes in a few microseconds.

Since the 1960s, many studies, using optical emission spectroscopy (OES) techniques, were realised for an experimental characterization of natural lightning. We can mention the important works of Uman *et al* [5] who deals with time-integrated spectra and Orville [6, 7] who performed the first time-resolved OES measurements on lightning. They obtained average temperature and electron density that reach, respectively, approximately 36000 K and $10^{18}$ cm$^{-3}$. More recent experimental studies on natural lightning strikes have also estimated the channel temperature and electron density using OES and under similar assumption and methods [8-10]. These authors reported average temperatures and electron densities that reach around 28000 K and $4 \times 10^{18}$ cm$^{-3}$, respectively.

The aim of the present work is to characterize high current pulsed laboratory arcs by the determination of several quantities such as the arc column and shock wave radius, and by performing space- and time-resolved determination of intensive thermodynamic properties as temperature, electron density and pressure. This work intent to extend and complete the characterization realized on previous studies on high current pulsed arcs ranging from 10 kA to 100 kA [11, 12]. A total of 11 waveforms, derived from three different current generators are used, covering current peak levels from 100 kA to 250 kA. These waveforms cover a wide range of current growth rates and action integrals (*AI*). This last parameter is obtained by integrating the square of current over time, with unit of $A^2$ s or J $\Omega^{-1}$.

The paper is organized as follows. In section 2, we present the current generators and the diagnostic setups utilized for high speed video imaging, Background-Oriented Schlieren (BOS) method and OES measurements. In section 3, we present the methods and procedures employed to characterize the arc. We describe the processing to obtained arc radius from luminous intensity profile and the wave front detection from BOS measurements. We present a briefly recall of the assumptions and the procedure to obtain a theoretical spectrum from the absorption coefficient calculation and by solving the radiative transfer equation (RTE). The results of column and shock wave radius, and radial profiles of temperature, electron density and pressure for all studied current waveforms are presented and discussed in section 4.

## 2. Experimental Setup

### *2.1 Current generators*

In this study we use three different current generators to produce high current pulsed arcs in a wide range of amplitudes, growth rates and action integrals. All studied current waveforms are derived from the criteria proposed by aeronautics standard and certification processes, as described in the document SAE ARP 5412A [4]. First characterizations of this kind of arcs were done using the GRIFON generator, which is a generator designed and developed at ONERA and is described in details in references [11, 13]. In a few words, the GRIFON generator is formed by four capacitors, connected in parallel resulting in a 200 µF capacitance, which discharge through a series of ballast resistors. The generator is designed in an approximate coaxial, where the current flows in the discharge located at the central axis of the structure and returns toward the capacitors by four symmetric aluminium bars. A self-inductance component appears from the structure and the switching is performed by a spark gap. Only one waveform of this generator is used in this work, with a peak of 100 kA.

The two others current generators used in this work are from DGA *Techniques aéronautiques* and cover a wide range of current amplitudes, and can provide current waveform with peak levels from 100 kA to 250 kA with different rise times and action integrals. Those are test benches normally used in certification processes for different aeronautical parts [4]. The SuperDICOM generator is a lightning high current generator used to perform important current levels (up to 350 kA) on high impedance samples. It is manufactured with a Marx generator which allows to deliver important voltage (up to 200 kV) through important internal resistance. It works as a RLC circuit with a critical damped response. The EMMA generator is also a RLC circuit but with a clamping system which permits to deliver up to 240 kA and to modify also the shape of the waveform by acting directly on the decay time. Both generators permit to deliver high currents by conduction or by creating an electrical arc on a sample which is connected to a test rig. The test rig is an aluminium plate with a pseudo-coaxial current return. A total of 11 current waveforms are studied in this work. Table 1 summarizes the main characteristics of the waveforms, and figure 1 shows the current measurements.

**Table 1**. Characteristics of analysed current waveforms used in this study.

| Generator | GRIFON | EMMA | | | | | | Super DICOM | | | |
|---|---|---|---|---|---|---|---|---|---|---|---|
| $I_{peak}$ (kA) | 100 | 94 | 97 | 147 | 149 | 200 | 240 | 100 | 150 | 200 | 250 |
| $t_{peak}$ (µs) | 13 | 18 | 25 | 25 | 28 | 28 | 34 | 14 | 19 | 21 | 24 |
| dI/dt (kA µs$^{-1}$) @ 2 µs | 14.5 | 10.9 | 6.3 | 12.3 | 11.5 | 14.7 | 14.2 | 14.2 | 16.9 | 20.2 | 21.8 |
| AI (MJ Ω$^{-1}$) | 0.28 | 0.25 | 0.29 | 0.66 | 0.95 | 1.7 | 3.35 | 0.25 | 0.6 | 1.8 | 2.5 |

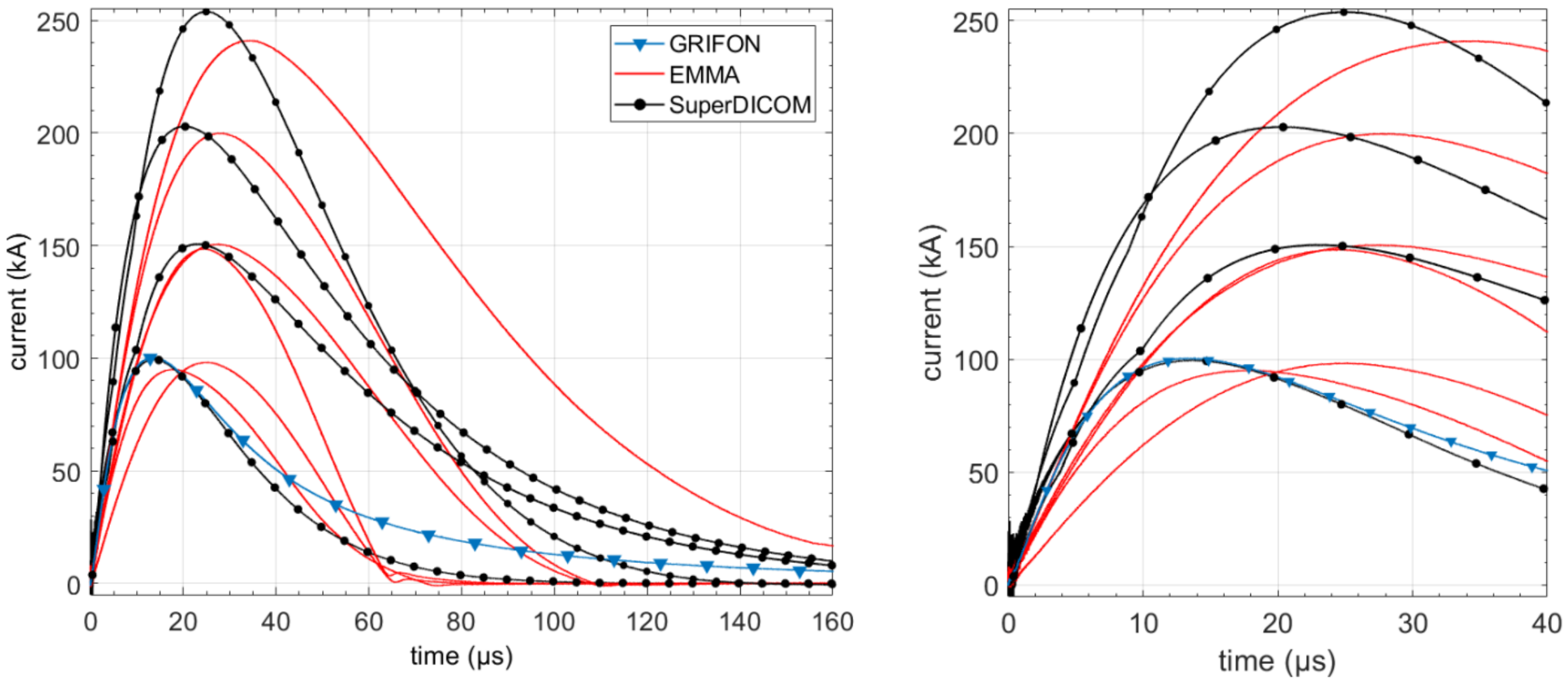

***Figure 1.*** *Current waveforms for different current generators and amplitudes. On the right side, we present a focus on the firsts 40 µs to highlight the current peak time and the growth rate.*

The electrode configuration is coherent with previous studies [11-13]. To allow the study of the electrical arc avoiding major effect from the electrode, they are formed by two identical tungsten rods with a dielectric sphere on the tip. This sphere is made of Araldite epoxy resin, and has a diameter of 20 mm. It is used to avoid a plasma jet directed to the cylindrical part of the arc channel, which would create instabilities on the free arc column and could also contaminate the air plasma with metallic species. Figure 2*(a)* illustrated the experimental setup and figure 2*(b)* shows a picture of the arc channel acquired by High-speed video camera (HSVC) where we can observe the two dielectric spheres and the cylindrical arc column formed on the central part. The distance between the two dielectric spheres is set to 120 mm. As a result of the ignition wire effect analysis performed at [11, 13], where was shown that carbon wires have less influence on the development of the arc column compared to copper ones, a carbon wire is chosen to breakdown the 120 mm gap and initiate the discharge.

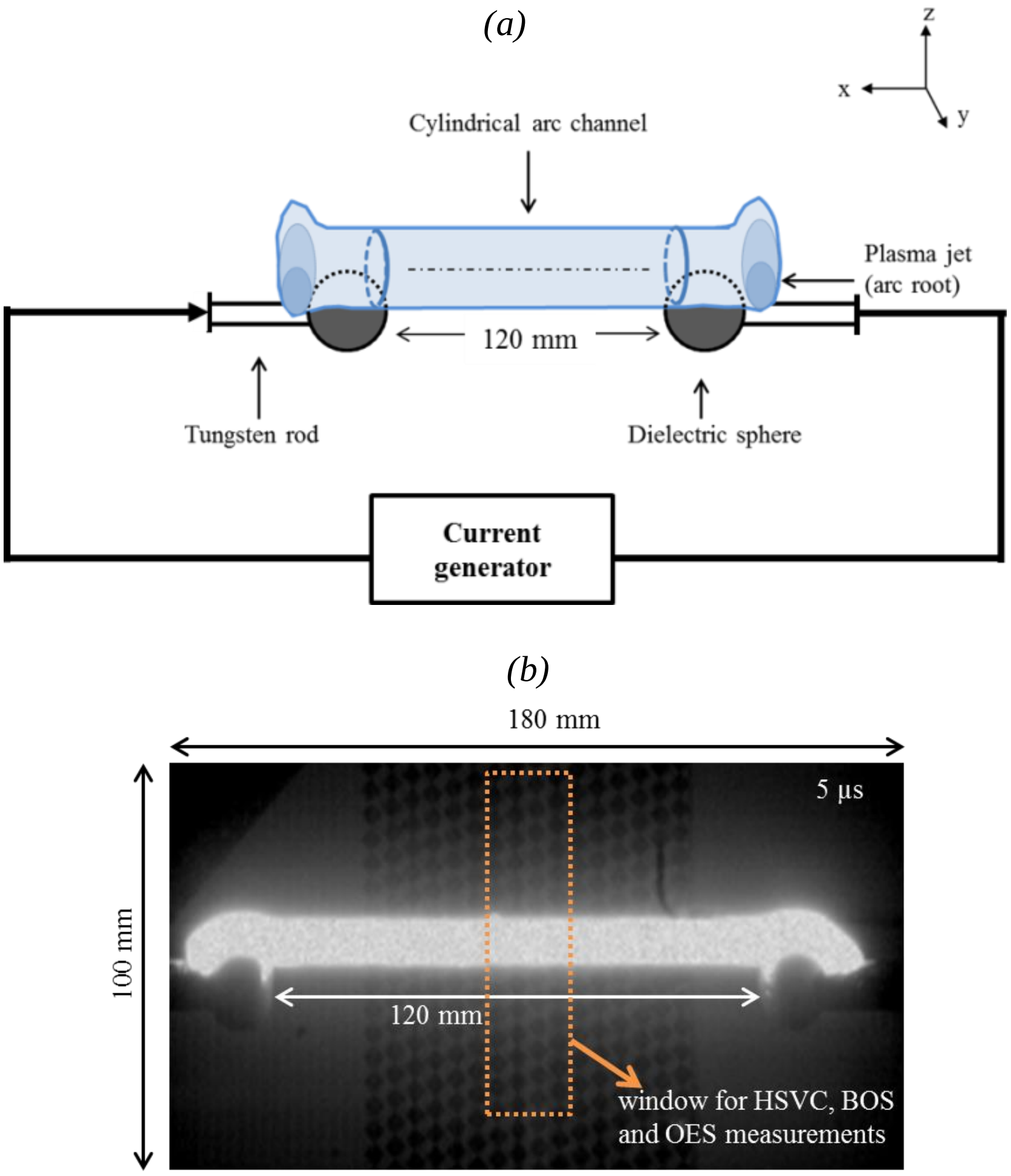

***Figure 2.*** *(a) Schematic for electrodes configuration. (b) Picture of the free arc column for the EMMA 200 kA waveform at 5 µs. The dotted rectangle indicates the zone for optical measurements of High-Speed Video Cameras (HSVC), Background-Oriented Schlieren (BOS) and Optical Emission Spectroscopy (OES). Resolution 0.70 mm/pixel, acquisition rate 79000 fps.*

### *2.2 Diagnostics setup*

In addition to electrical measurements as current and voltage probes, different optical diagnostics are employed to characterize the arc. In this work we use a few techniques already validated on previous studies [11-13] such as: 1. high speed video imagining for shape and radius evaluation of the channel; 2. Background-Oriented Schlieren (BOS) for shock wave analysis; and 3. Optical Emission Spectroscopy (OES) to perform a time- and space-resolved characterization of the intensive thermodynamic properties as temperature, electron density and pressure. In the following paragraphs we present a briefly description of the diagnostic setup. For more details on the setup, please refer to [11-13].

High speed video imaging and BOS is performed with two high speed video cameras (HSVCs) Phantom V711 from Vision Research (CMOS sensor of 1280 × 800 pixels of 20 µm$^2$), which can work up to 1.0 Mfps at reduced resolution. In this work, the resolution is set to 256 × 32 pixels, allowing a sample rate of 500 kfps. Exposure time is set to 300 ns and neutral-density filters (OD 3) are used to avoid saturation. The cameras are positioned in a perpendicular direction to the arc column axis and observe the central portion of the arc at mid-distance between the two electrodes. The observed area is indicated in figure 2*(b)*.

For OES measurements, an optical fibre bundle with 16 fibres (250 µm diameter for each fibre) disposed in a line array (500 µm between two fibre centres) is used to gather light from the arc to the spectrometer entrance slit. Two achromatic lenses (400 to 700 nm transmission range) are used to image a transverse plane of the arc column to the entrance of the fibres. The first lens is located at 1500 mm of the arc axis and has a focal length of 750 mm. The second one is placed at 350 mm from the first lens and has a focal length of 400 mm. The bundle entrance is located behind the second lens and the magnification factor of the optical system is adjusted to 4, which results in a collection of 16 chords spaced by 2 mm on the arc position. The spectrometer used is an ACTON SP-2750 with 750 mm focal length mounted with a grating of 300 grooves $mm^{-1}$ and an intensified CCD camera PI-MAX 2 Roper Scientific (512 × 512 pixels of 13 $\mu m^2$). The entrance slit aperture is set to 20 µm, which leads to a measured slit function with full width at half maximum (FWHM) of 0.166 nm. The acquired spectra are calibrated in relative intensity using a Deuterium Tungsten-Halogen Calibration Source from Ocean Optics.

## 3. Theory and methods

In this section we present the theory and procedures to analyse the data obtained by the three experimental diagnostics presented in section 2. The methods and data analysis used in this work were employed and described in details in previous studies [11, 12]. Therefore, in the following subsection we will briefly recall the different techniques and steps made to assess the quantities that characterize the arc channel.

### *3.1 Luminous intensity analysis*

The HSVCs used for luminous radius measurements acquire images of 100 × 12.5 $mm^2$ at the central part of the arc, on the *zy* plane on the reference frame of figure 2. This part of the arc channel develops as a cylindrical column in the first 50 µs. After this time, instabilities coming from the electrode's zone arise and the axisymmetric assumption is no longer valid [11]. We assume that the emitted light is a good criterion to evaluate the shape of the plasma. In order to define the radius based on the intensity profile measurements, two criteria are tested; one is based on the position of maximum rate of change of the intensity with radius and the second is based on a threshold of 20% of the maximum intensity of all images from a shot. Both criteria lead to approximately the same results, with a maximum difference of around 0.8 mm, which corresponds to roughly two pixels in the camera resolution (0.39 mm/pixel). The total uncertainties on the radius determination are calculated from the accumulation of statistical, criterion and instrumental errors, showing a minimum value of 0.9 mm at the first microseconds and a maximum of 2.4 mm around 50 µs.

### *3.2 Background-Oriented Schlieren*

After arc ignition, the fast transfer of a large amount of energy into a small volume increases the arc channel temperature, which leads to the channel expansion. This rapid expansion gives rise to a strong shock wave in the radial direction of the column. This shock wave exhibits a high gradient of the mass density at the wave front, modifying the local refractive index and making it possible the use of Schlieren techniques. As a result of the high radiative flux emitted by the arc, the BOS method is used to characterize this shock wave, which is a technique allowing the observation of refractive index gradient by the distortion of a patterned background [14, 15].

As indicated on figure 2*(b)*, the HSVC used for BOS measurements observe the arc column at the same position as the HSVC for radius measurements, with an image size of 171 × 21 $mm^2$. A patterned background formed by a black/white square mesh is used. The front wave position at each time is obtained by analysing the difference between the image to be analysed and a reference image taken without arc. The uncertainties associated with the shock wave radius determination are calculated as for the luminous radius, taking into account statistical and instrumental errors. The maximum observed total error is 1.6 mm.

### *3.3 Spectral lines and radiative transfer equation*

For OES measurements, we use a spectral window covering 523.3 to 576.6 nm. In prior studies, this spectral range was shown to be a convenient window, which has a significant intensity variation with temperature in the range of 20 000 K to 40 000 K, thanks to the N II multiplet located around 568 nm. In the present study we consider 31 lines of N II and 3 lines of O II for the calculation of the absorption coefficient. This coefficient is expressed as a function of temperature $T$ and electron density $N_e$ by assuming the arc column to be at local thermodynamic equilibrium (LTE) and by taking spectroscopic constants as central wavelength, energy upper levels, degeneracy and Einstein emission coefficient for each transition from NIST database [16] and total population of species calculated from HTGR database [17-19]. As a result of the high electron density, the spectral line shape is approximated by a Loretzian profile and the Stark broadening is assumed as the major effect in the line broadening, being considered proportional to the electron density and slightly dependent on the temperature [20, 21]. The Stark parameters as a function of $N_e$ are taken from [21].

In the previous work, it was shown that the optical thickness $\tau$ of arc column is very important for those levels of current and for the studied spectral range, with $\tau > 1$ for the major portion of the emission lines. Consequently, a procedure taking into account a non-optically thin medium is employed. Concisely, this procedure consists of split the arc column in $n$ circular and concentric layers, with $n$ corresponding to the number of collected chord of the optical system and the layer thickness being the distance between two adjacent chords (2 mm). The radiative transfer equation is then solved along the different chords, starting from the farthest chord that crosses just one layer until the central chord, which crosses all layers. For each chord, the calculated spectrum is convoluted with the slit function and then the result is compared with the measured data, both being normalized by their respectively highest value. A minimization by least square is performed to assess $T$ and $N_e$ for the corresponding layer. This procedure is similar to an Abel inversion, but taking into account self-absorption and avoiding spatial derivatives. At the end of this procedure, we are able to determine the profiles of $T$ and $N_e$. The pressure $P$ is then deduced from each couple ($T$, $N_e$) relying on LTE air plasma composition calculated in [17-19]. Figure 3*(a)* shows examples of spectra collected by each chord crossing the arc and figure 3*(b)* illustrates a fit of the measured data by a calculated spectrum, and the position of the considered ionic lines.

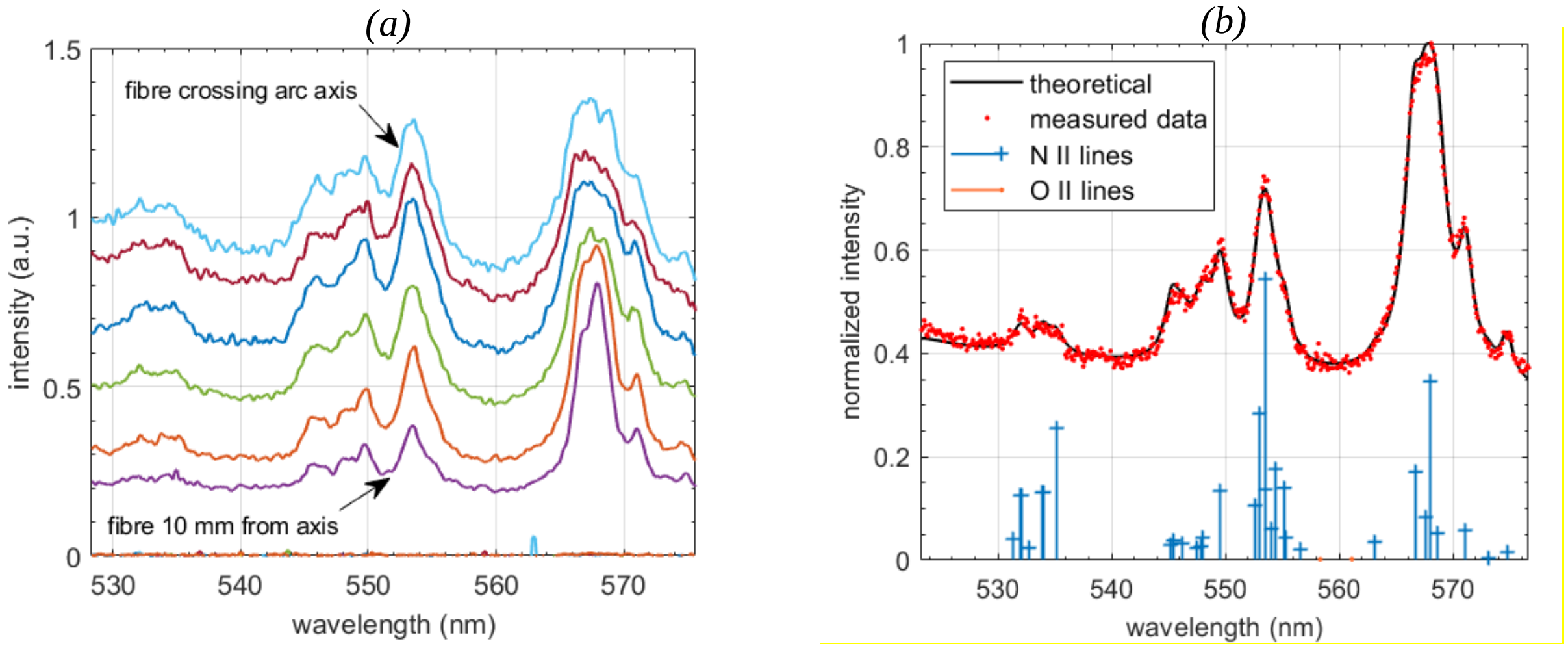

***Figure 3.*** *(a) Example of spectra collected by the diagnostic setup at different position inside the arc column. Experimental parameters: EMMA generator with 200 kA peak, time = 5 µs, exposure time = 200 ns. Each collected chord is spaced by 2 mm, with the less intense being at 10 mm from the arc axis. (b) Example of measured and calculated spectra to T and* $N_e$ *adjustment. The considered ionic line positions are presented, weighted by their corresponding degeneracy and Einstein emission coefficient. A multiplicative factor (*$10^{-9}$*) is applied to those lines to fit them in the normalized graph.*

The errors associated with the estimated parameters are difficult to obtain because the number of shots by configuration are very limited and then statistical analysis are not possible. Consequently, the criterion of parameter sensitivity used in [12], which is based on the root mean square error, is also applied here. From this criterion the errors that arise on temperature profile are around 12% at the column edge and reaches 22% on arc axis. The uncertainty on electron density reaches values of 38% and for the deduced pressure they are about 50%.

## 4. Results and discussion

In this section, we present the results of arc column radius, shock wave position and temperature, electron density and pressure distribution obtained using the methods described in section 3. For the sake of simplicity, the results are discussed as soon as they are presented.

### *4.1 Luminous radius measurements*

For the different waveforms studied in the present work, the axisymmetric assumption is verified to be valid for at least the firsts 50 µs, and then the luminous profiles are analysed on those times. Figure 4 shows the results for the waveforms from 100 kA to 250 kA peak. To an easy visualization and analysis, we arranged the curves of radius evolution on three plots and the error bars are not shown, however we remind that they range from 0.9 to 2.4 mm.

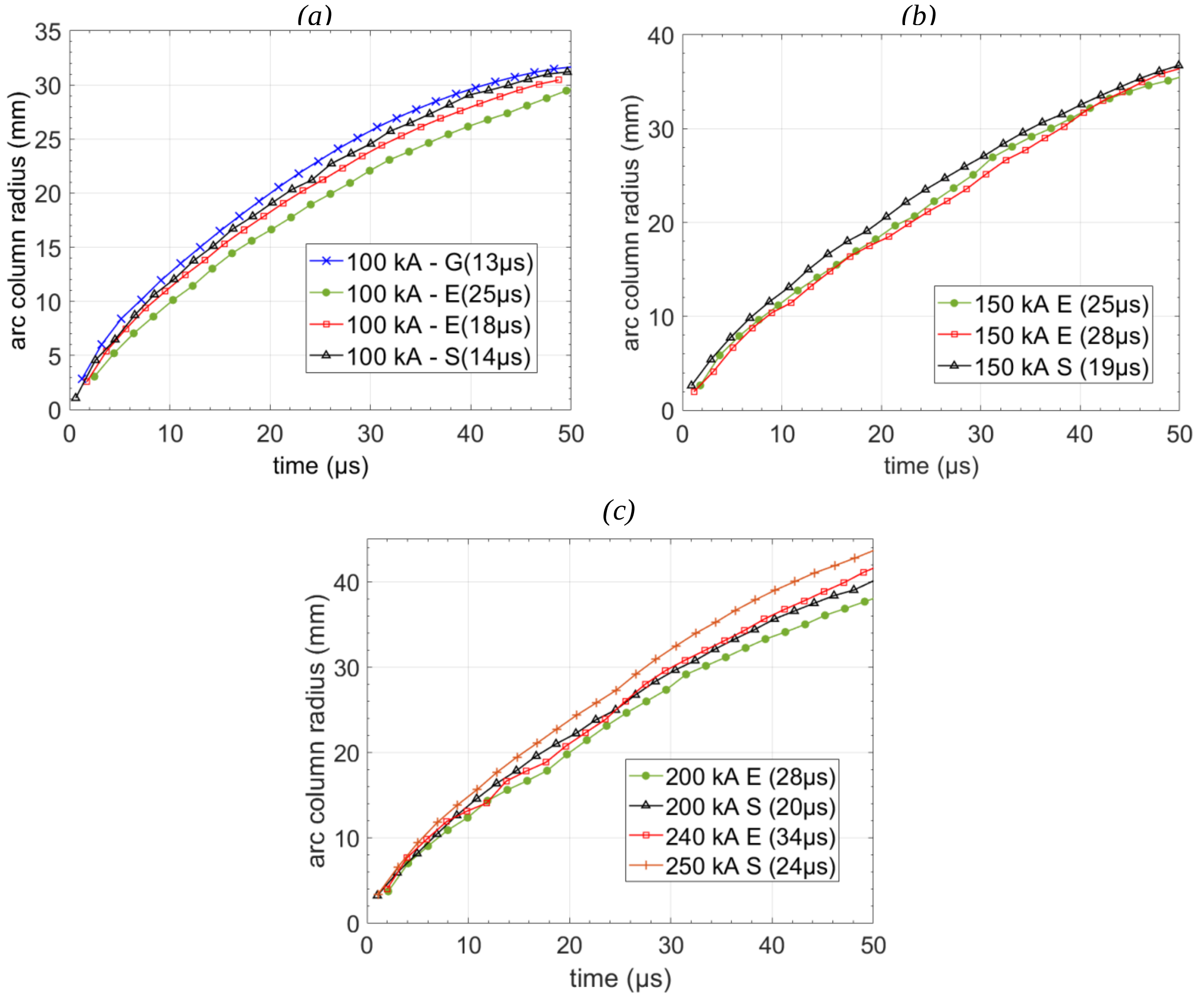

***Figure 4.*** *Luminous column radius comparison. (a) 100 kA; (b) 150 kA; (c) 200, 240 and 250 kA. On the legend the letters "G", "E" and "S "correspond respectively to the waveforms from the current generators GRIFON, EMMA and SuperDICOM. Times shown in parenthesis correspond to the current peak time of the associated waveform.*

In general, the arc radius increases quite monotonically with the current level and also with the reduction of the peak time, which for a same current level is equivalent to improve the current growth rate (*dI/dt*). We can observe in figure 4*(a)* that for 100 kA, a factor two on the *dI/dt* causes an increase of 20% of the column radius, reaching 32 mm for the faster waveform. The same behaviour is found on the waveforms of 150 kA, 200 kA and 250 kA, showing systematically higher radii for higher growth rates. The action integral (*AI)* seems to have a minor influence on the radius. Figure 4*(b)* shows the results for EMMA 150 kA waveform where a 50% higher *AI* (660 to 950 kJ $\Omega^{-1}$, corresponding respectively to E(25μs) and E(28μs)) leads to a difference of radius at 50 μs of less than 2 mm. For all current levels, the curve shape is similar and seems to follow a time root square law. Comparing the GRIFON waveform of 100 kA with the EMMA waveform of 240 kA, with both having approximately the same *dI/dt* (~14 kA $\mu s^{-1}$), we see that a factor 2.5 on the current amplitude result in a radius only 18% higher. Comparing the two extreme studied cases, the radius of the highest waveform (SuperDICOM 250 kA, 21.8 kA $\mu s^{-1}$), reaches 44 mm which is 57% higher than the slowest one (EMMA 100 kA, 6.3 kA $\mu s^{-1}$) that reaches 28 mm.

*4.2 Shock wave measurement*

It was shown on previous studies that the shock wave generated by the arc expansion has a cylindrical form, developing around the arc column and with a radial propagation [11, 13]. We focus our attention only on the central part of this cylindrical shock to obtain its radius over time. Figure 5 presents the results for all studied current waveforms. As for the case of the column radius, the results are presented on three plots, without error bars which have maximum value of 1.6 mm.

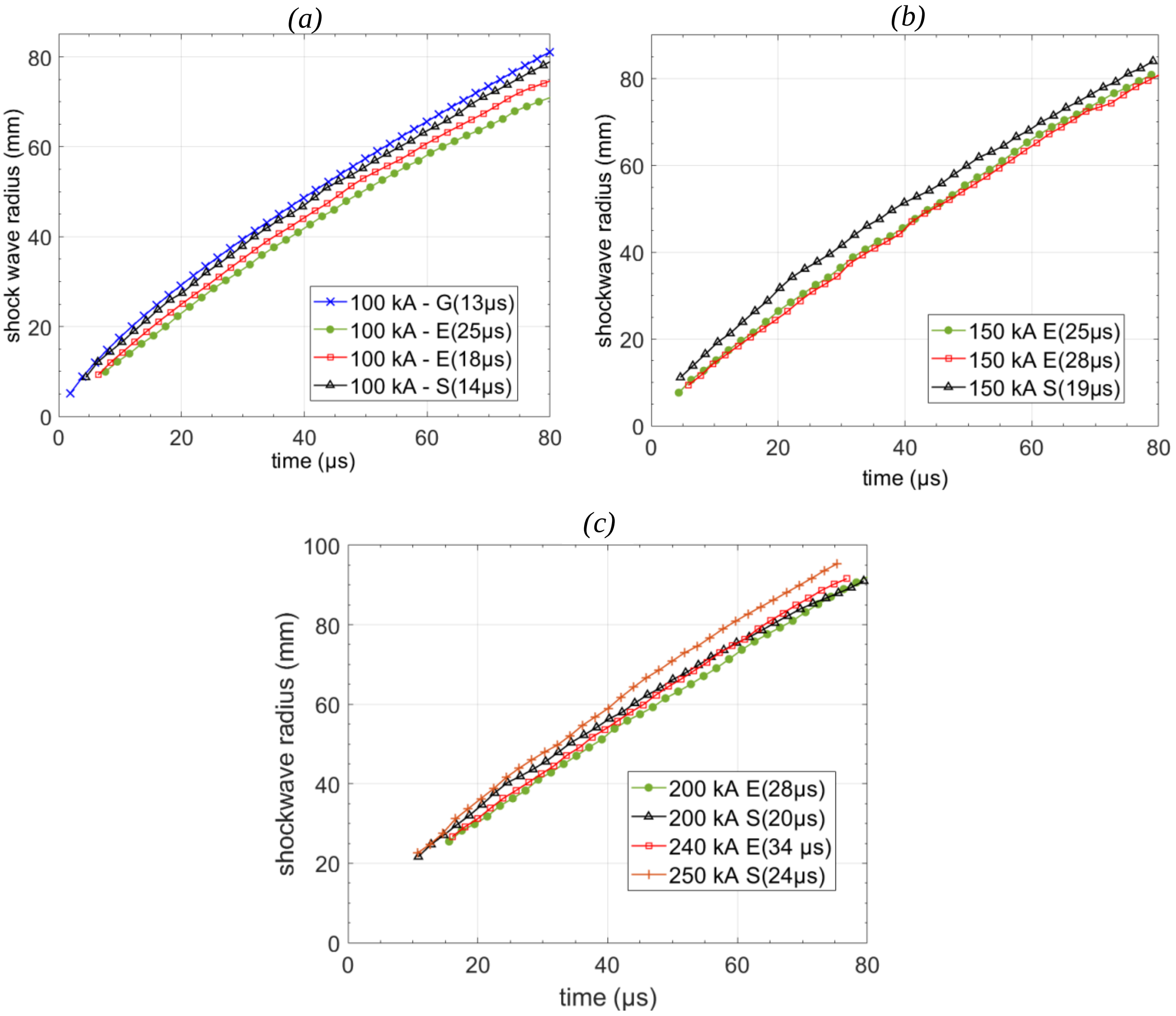

***Figure 5.*** *Shock wave comparison. (a) 100 kA; (b) 150 kA; (c) 200, 240 and 250 kA. On the legend the letters "G", "E" and "S "correspond respectively to the waveforms from the current generators GRIFON, EMMA and SuperDICOM. Times shown in parenthesis correspond to the current peak time of the associated waveform.*

The general rule found for the radius evolution is also observed for the shock wave. The shock radius increases with current peak and current growth rate. However, the shape over time is more linear for the shock radius when compared to the arc radius. Actually, the shock wave losses energy on the shocked gas, but propagation velocity decreases slower than the arc radius evolution, which reaches some kind of saturation and stop growing.

It is difficult to separate and quantify in a simple way the influence of the current peak and current growth rate on the shock wave. It can be seen, for example, that shocks from slow waveforms at 150 kA (EMMA with 25 µs and 28 µs of peak time) are less than or equal to those of faster waveform at 100 kA (GRIFON with 13 µs of peak time). The effect of *AI* on shock waves seems to be negligible. A 50% variation on *AI* for the 150 kA waveform does not modify the shock propagation. This result is coherent, since for current waveforms with the same peak value and almost the same *dI/dt*, the instant when the current levels differ occurs after several microseconds, between 30 to 50 µs for those waveforms. At these times, the shock is already detached from the arc column and is more than 10 mm far from it, being no longer influenced by the injection of the current. Shock measurements are performed up to 93 mm due to the acquisition window, however the cylindrical shock propagates until it reaches the generator current return structure. For the highest current waveform at 250 kA (SuperDICOM 250 kA, 21.8 kA $\mu s^{-1}$), the shock radius reaches 93 mm at 74 µs, while at the same time the slowest waveform (EMMA 100 kA, 6.3 kA $\mu s^{-1}$) is 30% lower, with a value of 67 mm. At the same current level, a factor two on the *dI/dt* causes an increase of around 20% on the shock wave radius (comparison at 100 kA for EMMA and GRIFON waveform).

### *4.3 Temperature, electron density and pressure measurements*

From the OES measurements and by solving the RTE for different chord as described in section 3, we assess the profiles of temperature, electron density, and, from those quantities, we deduce the pressure distribution. Aiming to achieve a space- and time-resolved characterization, for each current waveform, the measurements are performed at four instants; 6, 9, 14 and 25 µs. Depending on the waveform the last time varies to 24 or 26 µs.

The physical quantities analysed are influenced by the current peak level but also by the current growth rate. In general, as for the case of column and shock wave radius, the properties increase with the increase of these two parameters. For a given instant, the temperature distribution within the arc is approximately constant and presents an important gradient at the edge of the column. This homogenization of the temperature in the column for those current levels is related to radiative effects [19, 22]. The radiative emission transports energy from the hotter region on the centre arc to the periphery as a result of the absorption of the emitted intensity by the surrounding cold gas. This mechanism contributes also to the expansion of the column radius. The electron density distribution has an approximately parabolic shape, with a maximum value in the axis and drops along the radial direction. This behaviour is related to Laplace/Lorenz forces, due to the interaction between the strong current flowing in the arc and the self-magnetic field derived from this current [23, 24]. This force is radial and directed towards the centre of the column and decreases with $r^2$. The pressure profiles, deduced from the pair ($T$, $N_e$), have the same shape as that of the electron density profiles. At the analysed times, the obtained profiles have their maximum values at 6 µs and decay over time. This spatial-temporal behaviour for $T$, $N_e$, and $P$ is consistent with previous studies for high pulsed current arc up to 100 kA [13].

For the sake of clarity, the determined profiles are presented in separated graphs, sorted by current levels. For each profile, the farthest radial position is set to the standard ambient conditions (298.15 K, 1 bar), since no intensity was detected at these positions. We recall that the uncertainties associated with the results are estimated at 22% for temperature, 38% for electron density and can reach around 50% for pressure.

Figure 6, 7 and 8 shows results of respectively, temperature, electron density and pressure for the four current waveforms of 100 kA peak. For all quantities, the maximum value is reached at the first measurement instant (6 µs) and then decreases with time. The maximum temperature reached is around 33000 K. A factor of about two on *dI/dt* (comparison between EMMA and GRIFON

waveform) gives rise to a change of 12% in temperature (a difference of about 4000 K). The electron density and the pressure on the other hand are more impacted by the *dI/dt*, changing from 26.4 to 39.8 × $10^{17}$ $cm^{-3}$ and 21.5 to 35 bar respectively, which corresponds to about 35%.

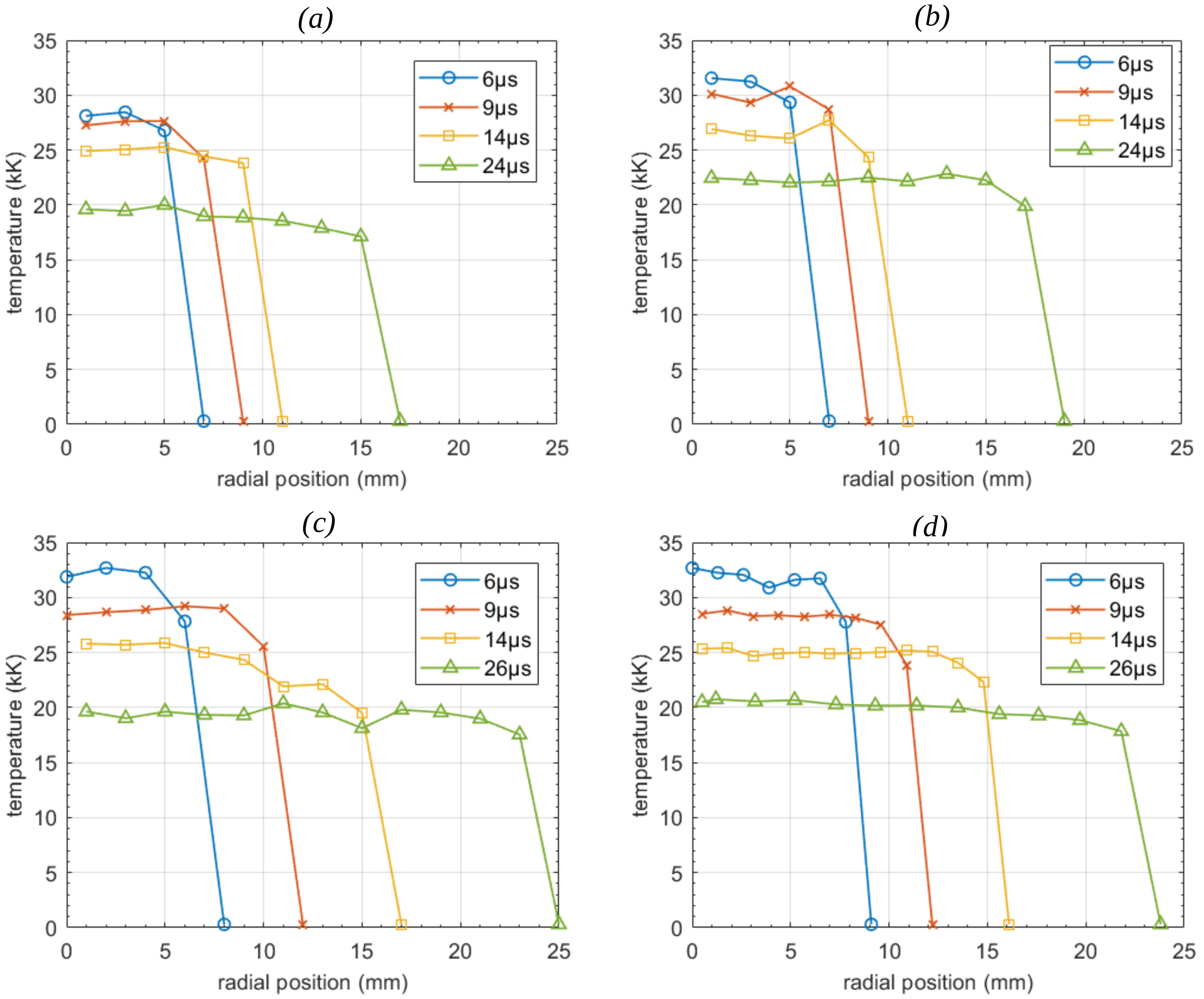

***Figure 6.*** *Temperature distribution for different waveforms with 100 kA peak. (a) EMMA 100kA slow (tp =25μs), (b) EMMA 100kA fast (tp = 18μs), (c) SuperDICOM 100 kA (tp = 14μs), (d) GRIFON 100kA (tp = 13 μs)*

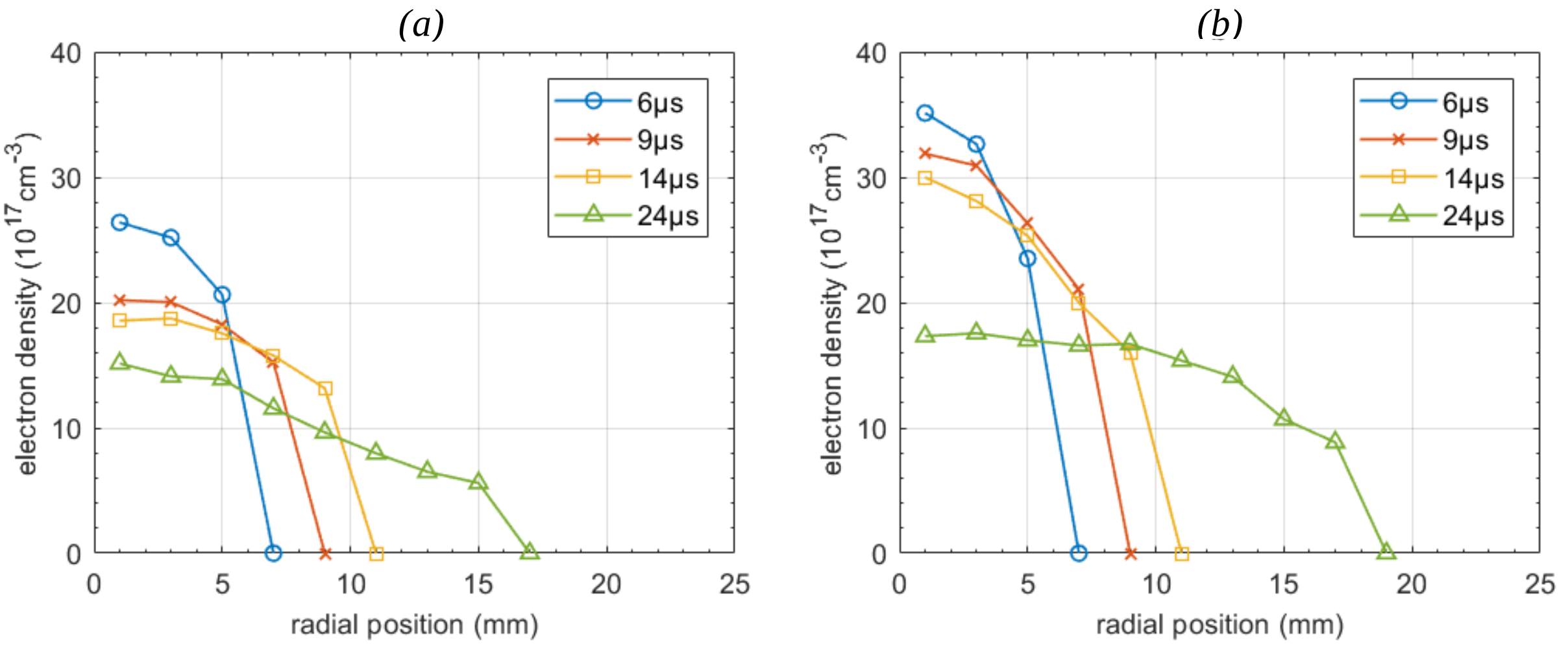

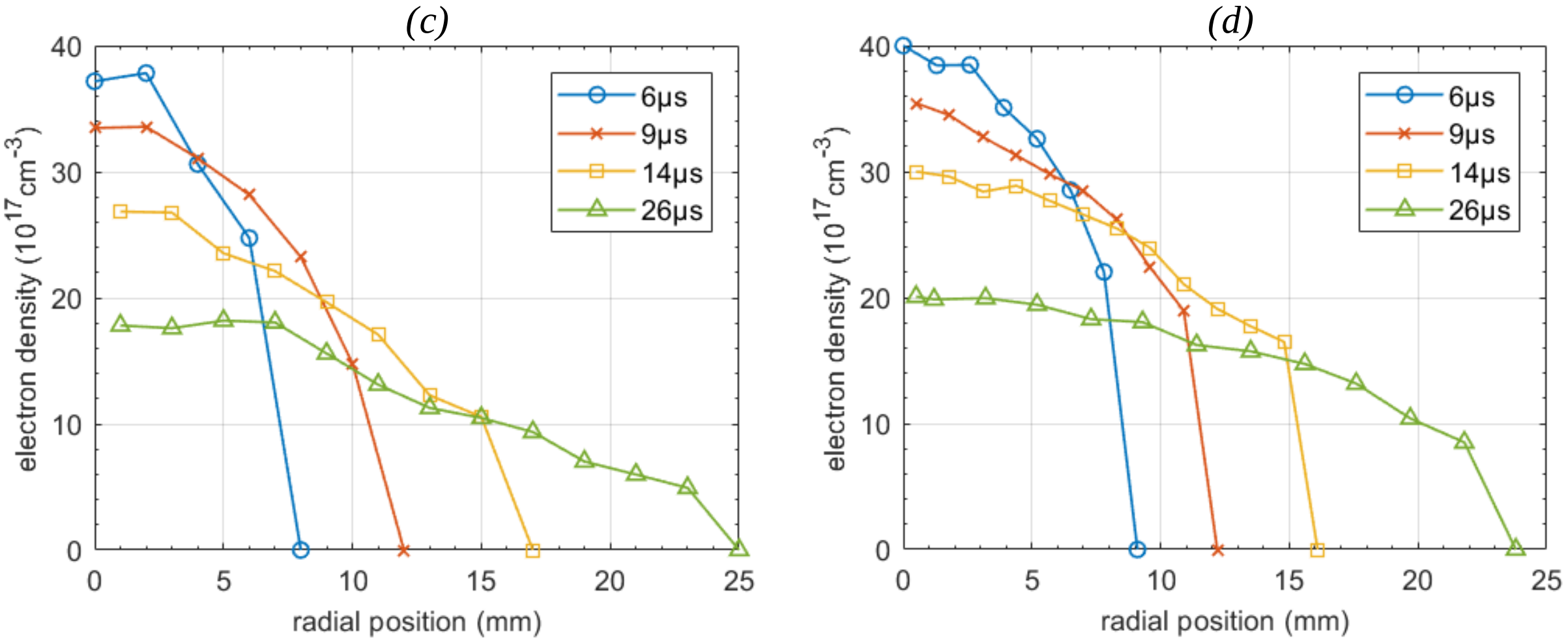

***Figure 7.*** *Electron density distribution for different waveforms with 100 kA peak.*
*(a) EMMA 100kA slow (tp =25μs), (b) EMMA 100kA fast (tp = 18μs),*
*(c) SuperDICOM 100 kA (tp = 14μs), (d) GRIFON 100kA (tp = 13 μs)*

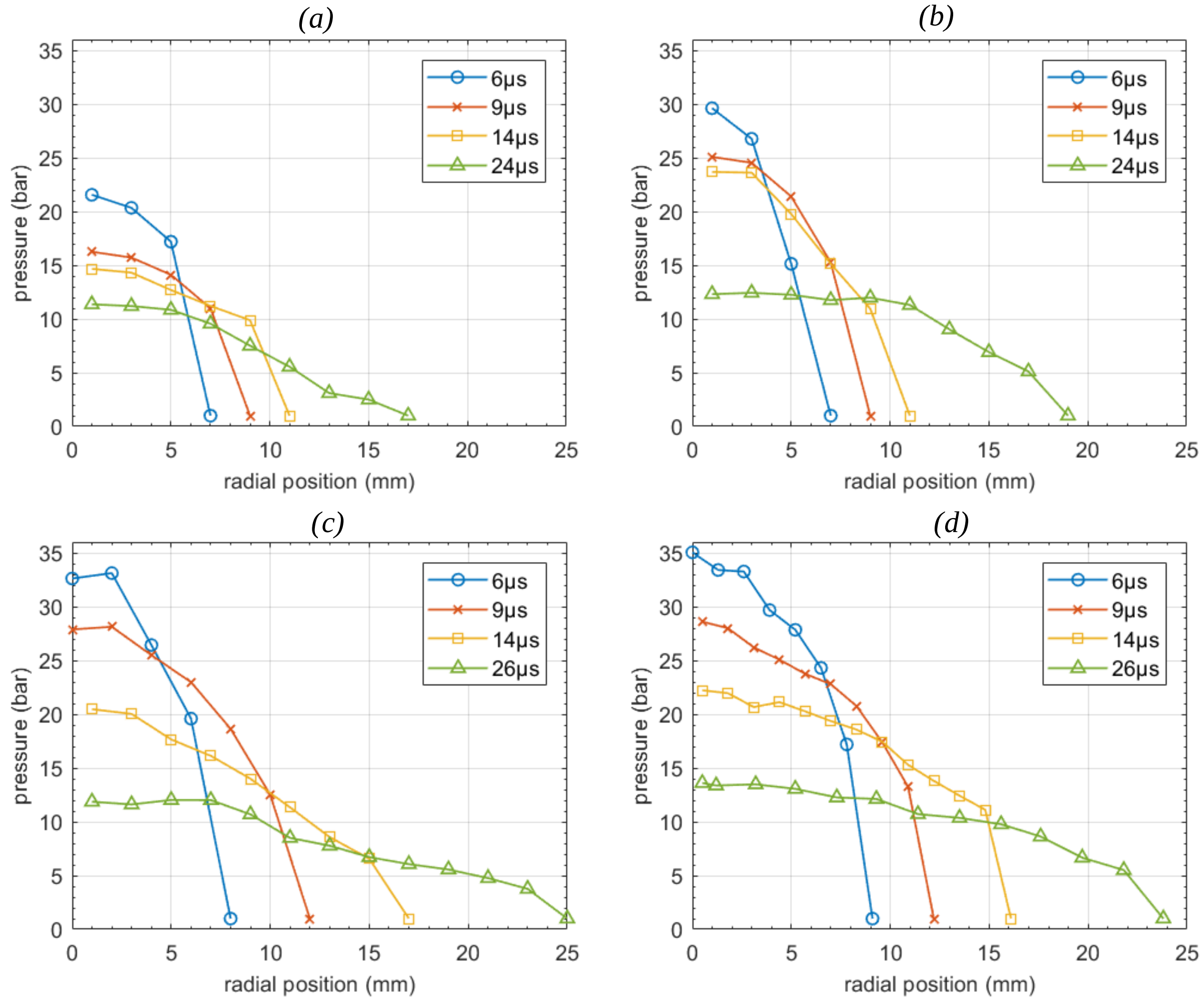

***Figure 8.*** *Pressure distribution for different waveforms with 100 kA peak.*
*(a) EMMA 100kA slow (tp =25μs), (b) EMMA 100kA fast (tp = 18μs),*
*(c) SuperDICOM 100 kA (tp = 14μs), (d) GRIFON 100kA (tp = 13 μs)*

Figure 9 presents the results for the two waveforms at 150 kA, from EMMA and SuperDICOM. Since in the analysed times, there is quite no difference on the current level for the two EMMA waveforms, we kept only that with 25 µs of peak time. As for the case of 100 kA, the higher values are also obtained at 6 µs. The temperatures increase by about 2000 K (6%) when compared to the 100 kA waveforms reaching 35000 K. The electronic density remains almost the same as for 100 kA. This can be explained by the increase of the column radius in the case of 150 kA, which promotes a decrease in density and pressure. For these waveforms, once the *dI/dt* are close (difference of around 5%), the physical properties are not very different between EMMA and SuperDICOM.

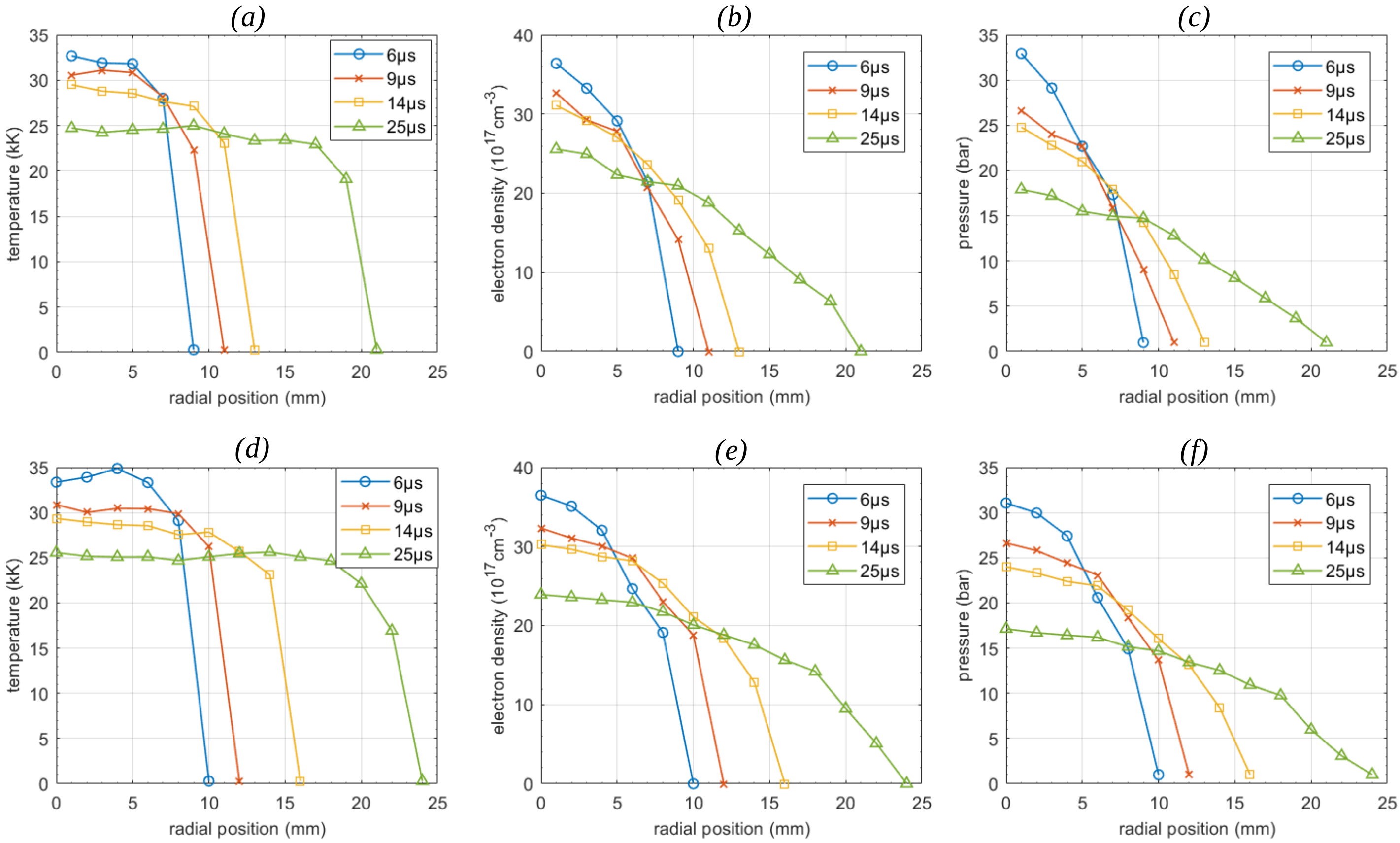

***Figure 9.*** *Temperature, electron density and pressure distribution for EMMA 150 kA (25µs) waveform (a-c) and SuperDICOM 150 kA waveform (d-f)*

Figure 10 presents the results for the two waveforms at 200 kA. The density and pressure increased by approximately 20% when compared to 150 kA. They reach values of the order of 46 × $10^{17}$ cm$^{-3}$ and 40 bar for the case of SuperDICOM. The maximum temperature is 39000 K, and the difference of 25% of *dI/dt* causes a difference of less than 10% on all properties.

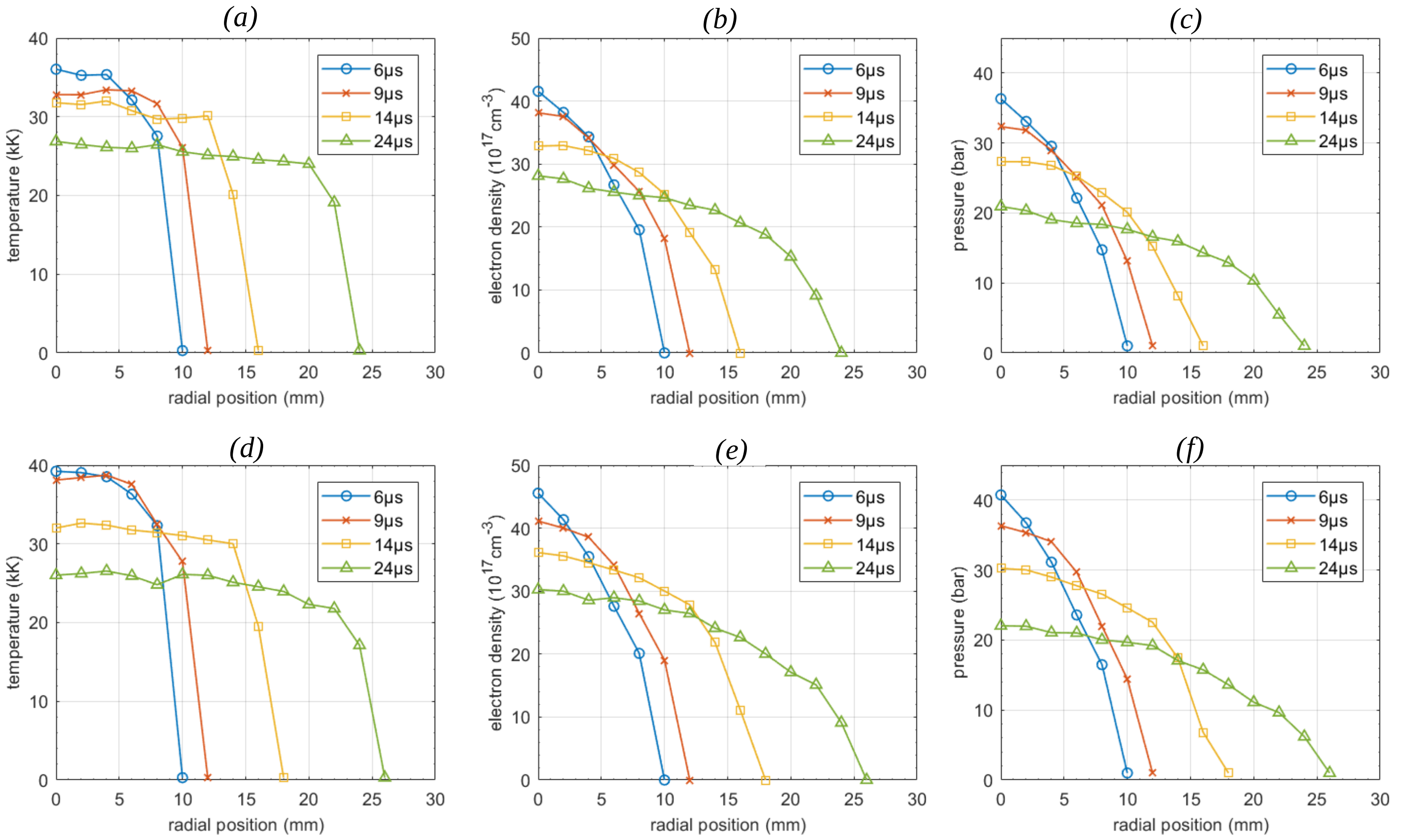

***Figure 10.*** *Temperature, electron density and pressure distribution for EMMA 200 kA waveform (a-c) and SuperDICOM 200 kA waveform (d-f).*

Figure 11 presents the results for the waveforms at 240kA and 250 kA. The behaviour of the properties follows that one as for lower current levels. A small increase of $T$, $N_e$ and $P$ is noticed (~10%) when we compare the two available *dI/dt* (~30% difference). The maximum values reaches at 6 μs are respectively 42400 K, 49.8 × $10^{17}$ cm$^{-3}$ and 45.9 bar. At 24 μs those values decreases around 30% for both waveforms.

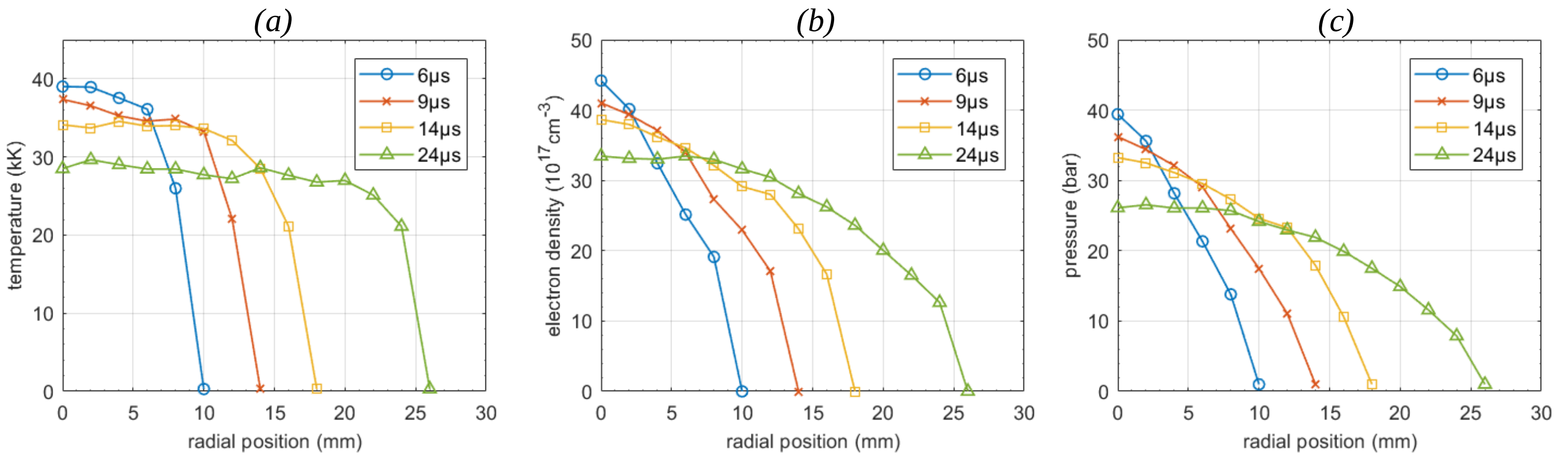

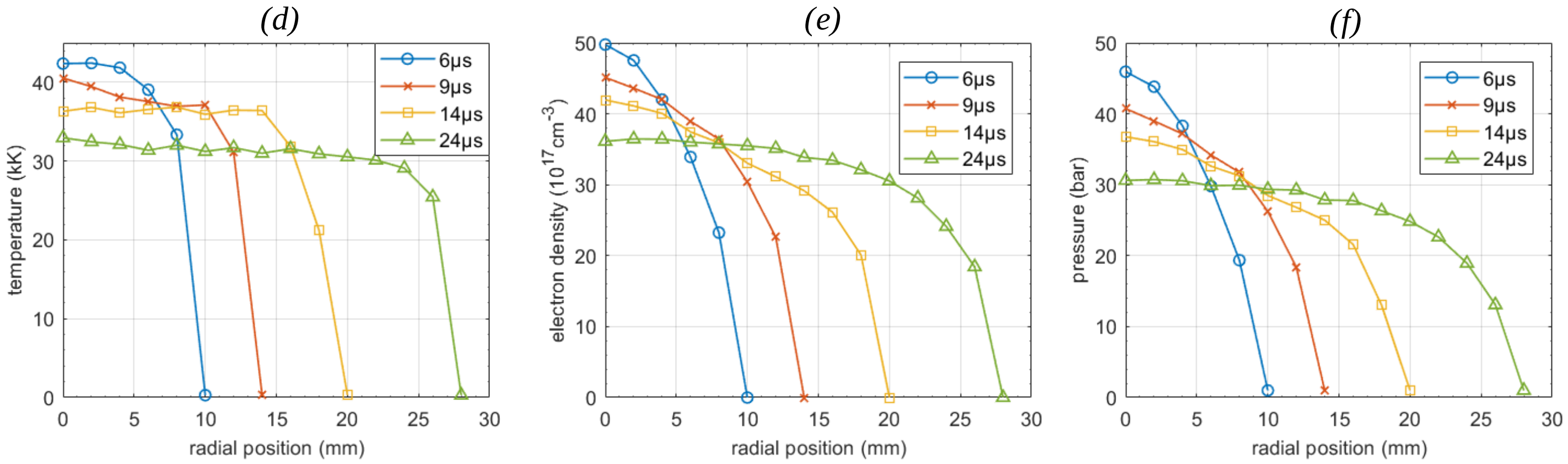

***Figure 11.*** *Temperature, electron density and pressure distribution for EMMA 240 kA waveform (a-c) and SuperDICOM 250 kA waveform (d-f)*

For a direct evaluation of the impact of the current peak level on the thermodynamic properties, table 2, 3 and 4 summarizes the results, listing the maximum reached value at each time. We observe that a factor 2.5 on the current level increases *T* of 30% and *P* of around 40% on the firsts microseconds. At 26 µs, the difference increases to 60% and 200% respectively.

Overall, the intensive thermodynamic properties have a moderate increase with respect to the applied current level, but it is important to point out that despite a small difference between these absolute values, the arc column radii are also growing with the increase of current and *dI/dt*. This implies a spread of those quantities over a larger surface, and consequently contributes to the establishment of some kind of balance between properties and current levels.

***Table 2***. *Maximum temperature (kK) reached inside the arc column for different instants and current peak levels.*

| | **100 kA** | | | | **150 kA** | | **200 kA** | | **250 kA** | |
|---|---|---|---|---|---|---|---|---|---|---|
| time (µs) | EMMA @25µs | EMMA @18µs | SDICOM @14µs | GRIFON @13µs | EMMA @25µs | SDICOM @19µs | EMMA @28µs | SDICOM @21µs | EMMA @34µs | SDICOM @24µs |
| 6 | 28.5 | 31.6 | 32.7 | 32.7 | 32.7 | 34.9 | 36.1 | 39.2 | 39.0 | 42.4 |
| 9 | 27.6 | 30.8 | 29.2 | 28.8 | 31.1 | 30.9 | 33.4 | 38.7 | 37.4 | 40.5 |
| 14 | 25.3 | 27.7 | 25.9 | 25.4 | 29.5 | 29.4 | 32.0 | 32.6 | 34.6 | 36.8 |
| 24/26 | 20.0 | 22.8 | 20.4 | 20.8 | 25.0 | 26.7 | 26.9 | 26.5 | 29.7 | 32.9 |

***Table 3**. Maximum electron density (*$10^{17}$ $cm^{-3}$*) reached inside the arc column for different instants and current peak levels.*

| | 100 kA | | | | 150 kA | | 200 kA | | 250 kA | |
|---|---|---|---|---|---|---|---|---|---|---|
| time (µs) | EMMA @25µs | EMMA @18µs | SDICOM @14µs | GRIFON @13µs | EMMA @25µs | SDICOM @19µs | EMMA @28µs | SDICOM @21µs | EMMA @34µs | SDICOM @24µs |
| 6 | 26.4 | 35.1 | 37.8 | 39.8 | 36.4 | 36.5 | 41.5 | 45.5 | 44.2 | 49.8 |
| 9 | 20.2 | 31.9 | 33.6 | 35.4 | 32.7 | 32.3 | 38.1 | 41.1 | 41.0 | 45.1 |
| 14 | 18.7 | 30.0 | 26.8 | 29.9 | 31.1 | 30.2 | 32.9 | 36.1 | 38.6 | 41.9 |
| 24/26 | 15.1 | 17.6 | 18.2 | 20.1 | 25.6 | 23.9 | 28.2 | 30.2 | 33.5 | 36.5 |

***Table 4**. Maximum pressure (bar) reached inside the arc column for different instants and current peak levels.*

| | 100 kA | | | | 150 kA | | 200 kA | | 250 kA | |
|---|---|---|---|---|---|---|---|---|---|---|
| time (µs) | EMMA @25µs | EMMA @18µs | SDICOM @14µs | GRIFON @13µs | EMMA @25µs | SDICOM @19µs | EMMA @28µs | SDICOM @21µs | EMMA @34µs | SDICOM @24µs |
| 6 | 21.5 | 29.6 | 33.1 | 35.0 | 30.8 | 31.5 | 36.3 | 40.7 | 39.4 | 45.9 |
| 9 | 16.3 | 25.1 | 28.1 | 28.6 | 26.6 | 26.6 | 32.4 | 36.3 | 36.2 | 40.7 |
| 14 | 14.7 | 23.7 | 20.5 | 22.2 | 24.8 | 24.0 | 27.3 | 30.2 | 33.2 | 36.8 |
| 24/26 | 11.4 | 12.4 | 12.0 | 13.6 | 19.9 | 17.1 | 20.9 | 22.1 | 26.5 | 30.7 |

## 5. Conclusion

In this work we performed an experimental investigation of the hydrodynamic and intensive thermodynamic properties of high current pulsed arcs. The studied current waveforms cover a wide range of amplitudes, extending from 100 to 250 kA, but also of current growth rates and action integrals. The present studied complete the previous ones that deal with current levels from 10 kA to 100 kA.

High speed imaging and the BOS method were used to assess, respectively, the temporal dynamics of the column arc and the associated shock wave. Both quantities are impacted by the current peak level and *dI/dt*. A cylindrical expansion of the arc column is observed for times up to 50 µs. The shock wave also propagates in a cylindrical shape around the arc channel. At 50 µs, the column radius reaches 28 mm for the slower 100 kA waveform and is 57% higher for the faster 250 kA, reaching 44 mm. For same peak levels, a factor two on the *dI/dt* leads to an increase of 20% of the radius. The same behaviour is observed for the shock wave evolution. The shock wave radius for the highest current waveform (at 250 kA) reaches 93 mm at 74 µs, while at the same time the slowest waveform (at 100 kA) is about 30% lower, with a value of 67 mm. Action integral has a minor effect on those quantities once for the analysed times the current waveform are similar.

Temperature, electron density and pressure profiles are determined by OES measurements, associated with a resolution of the radiative transfer equation under a non-optically thin medium assumption. As observed in previous works, temperature distributions show an approximately constant value within the arc column, and the electron density and pressure present roughly a parabolic shape along the radial direction. As for the case of column and shock wave radius, the properties are influenced both by current peak and *dI/dt*. At 6 µs, the temperature reaches 42400 K for the faster 250 kA waveform and it is 50% inferior for the slower 100 kA one. At the same current amplitude, a factor two on *dI/dt* leads to an increase of 12% of the temperature and around 35% for electron density and pressure. Since the column radius is also growing with the increase of current and *dI/dt*, the properties are spread over a larger surface, contributing to a balance within the column and consequently the impact of these parameters are moderated.

These results presented in this paper will serve to extend a database for simulation codes for this category of high pulsed arcs, and will be very useful and for model development and for validations of computational tools.

**Acknowledgment**

The authors wish to thank the French Civil Aviation Authority (DGAC) for its support.